\documentclass[preprint,a4paper,doublespacing]{revtex4}
\usepackage[dvips]{graphicx}

\begin{document}
\title{The interface between a polar perovskite oxide and silicon from monoatomic lines}
\author{I. Devos}
\email{Isabelle.Devos@isen.fr}
\affiliation{Institut d'Electronique, de Micro\'electronique et de Nanotechnologie, Unit\'e Mixte de Recherche CNRS 8520, Avenue Poincar\'e BP 69, F-59652 Villeneuve d'Ascq CEDEX FRANCE}
\author{P. Boulenc}
\affiliation{ST Microelectronics, 
850 rue Jean Monnet, F-38926 Crolles CEDEX FRANCE}
\date{\today}

\begin{abstract}
We report a study on the interface between polar high-$\kappa$ materials and the Si(001)-(2$\times$1) reconstructed surface with LaAlO$_3$ taken as a prototype material. The construction of the interface is based on the prior growth of metal lines followed by oxidation, whose stability against oxygen coverage is studied. Electronic structure calculations within the Density Functional Theory framework help in building the interface and understanding its bonding structure. Moreover, we computed a conduction band offset of 1.9~eV, in agreement with electronic applications requirement. The results may provide a guidance for interface processing.
\end{abstract}
\pacs{68.35.-p, 71.15.Mb, 71.15.Nc, 73.20.At}
\maketitle

The search for alternative dielectrics to silicon dioxide in MOS capacitors has recently focused on complex oxide materials with high dielectric constant (high-$\kappa$ materials). Many of them belong to the class of perovskites (cubic crystals whose chemical formula is ABO$_3$) which are thought to have a good ability to grow as crystalline films. However, the interface with Si is of crucial importance as it must avoid gap states and be structurally compatible with the film's crystallographic structure.
In the search of the best candidate, the LaAlO$_3$ crystal raised out of the perovskite family. This was partly due to its lower mismatch with the Si(001) surface (0.7~\%). Moreover, the conduction band offset is expected to be greater than 1~eV. A value has been computed\cite{RobertsonLAO} to be 1~eV within the "charge neutrality level model" which does not take the interface structure into account. Nevertheless, as La, Al and O respective oxidation states are +III, +III and -II, the AlO$_2$ and LaO layers are respectively charged with -1 and +1 electron which makes it a polar crystal. This polar aspect leads to a more complex interface construction than dealing with non polar perovskites.

This letter focuses on the interface between a polar perovskite and Si(100)-p(2$\times$1), taking LaAlO$_3$ as a prototype. The interface may start with the deposition of La or Al atoms. For an interface starting with La, Ashman et al. \cite{AshmanLa1} have shown that La atoms cover the Si surface by pairs. This scheme is coherent with the "electron counting model" where 2 La atoms provide 6 electrons to 6 Si dangling bonds, removing gap states and assuming a coverage of $2/3$ monolayer (ML). Klenov et al. \cite{Klenov} have grown a Si thin film on a LaAlO$_3$ substrate and microscopy has shown that the interface was based on La pairs. This interface has been theoretically studied by F\" orst et al. \cite{ForstLa2} to define the interface's Si and O networks that can not be resolved by the microscopy. The Si network obtained by Si epitaxy is not clearly coherent with LaAlO$_3$ epitaxy on the (2$\times$1) reconstructed Si surface. In order to avoid dipole moment of the polar surfaces of LaAlO$_3$, Knizhnik et al. \cite{Knizhnik} have constructed theoretical interfaces by transfering O from one boundary surface to the other in their LaAlO$_3$ slab.

Our aim here is to build an interface starting with Al atom deposition, compatible with the growth of LaO-AlO$_2$ neutral bilayers, without midgap states. 

\begin{figure}\centering
\includegraphics[width=6.cm]{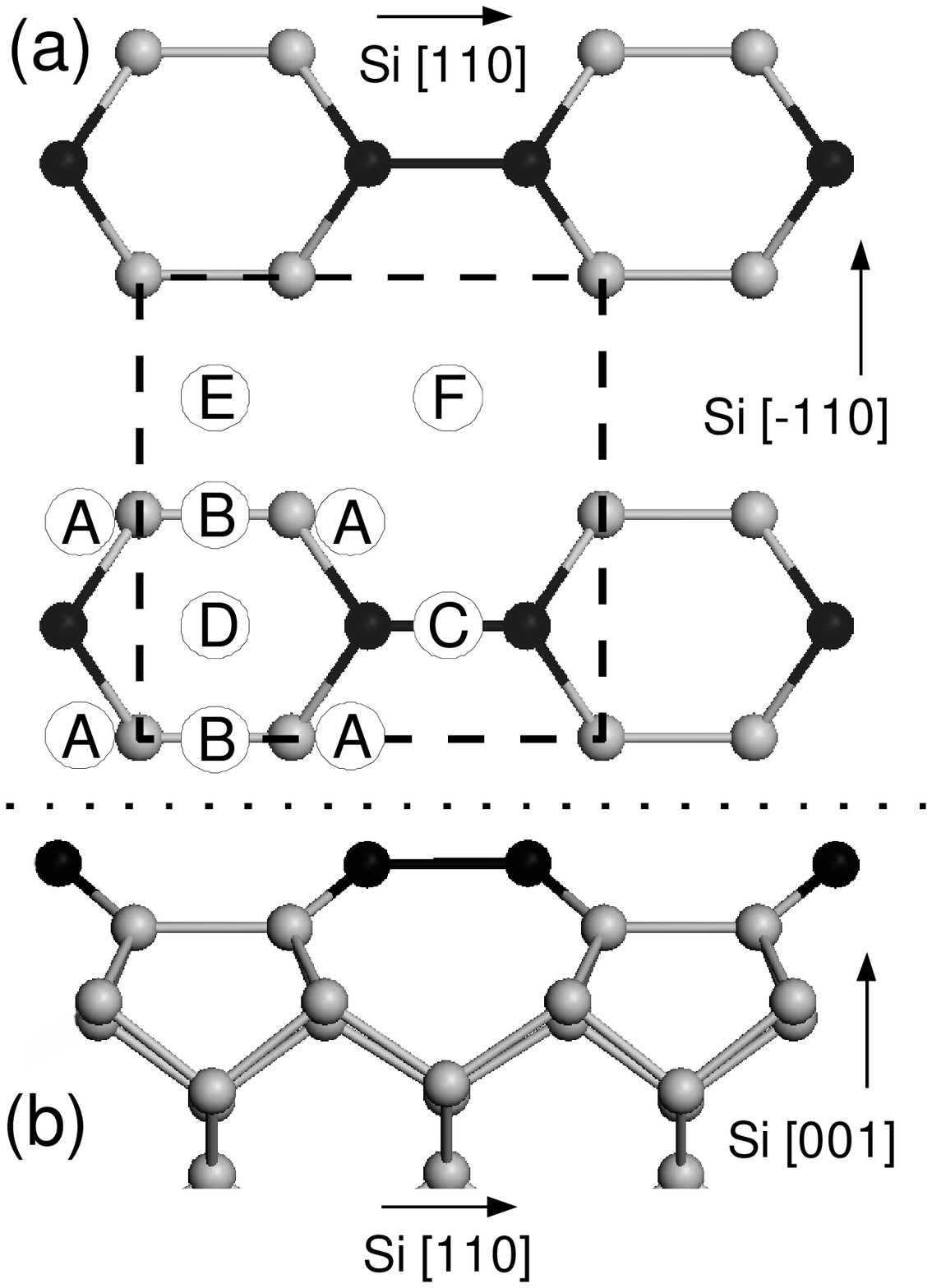}
\caption{}
\label{fig1}
\end{figure}

Due to the odd oxidation state of Al (+III), atoms are arranged by pairs on the Si surface. Indeed, it is well known for indium, aluminum and so on that the most stable geometry (Fig.\ref{fig1}) is realized at a coverage of $1/2$~ML by pairs of adatoms forming monoatomic chains \cite{Allines}. As the coverage of Si by Al lines has been realized ten years ago, we take this surface as a starting point for the growth of LaAlO$_3$.

The electronic structures were computed thanks to the SIESTA code \cite{Siesta}. We made use of the same pseudopotentials and basis sets as Knizhnik et al. \cite{Knizhnik}. The atomic basis for La, Al and O atoms were single-zeta plus polarization, and Si basis were double-zeta plus polarization. The supercells were fully relaxed in the Harris functional framework (until a maximun force tolerance of 0.1~eV/$\AA$ is reached). Then the electronic structure of the relaxed supercell was calculated in the Self-Consistent-Field framework, using a 2$\times$2$\times$1 k-point grid.

\begin{figure}\centering
\includegraphics[width=6.cm]{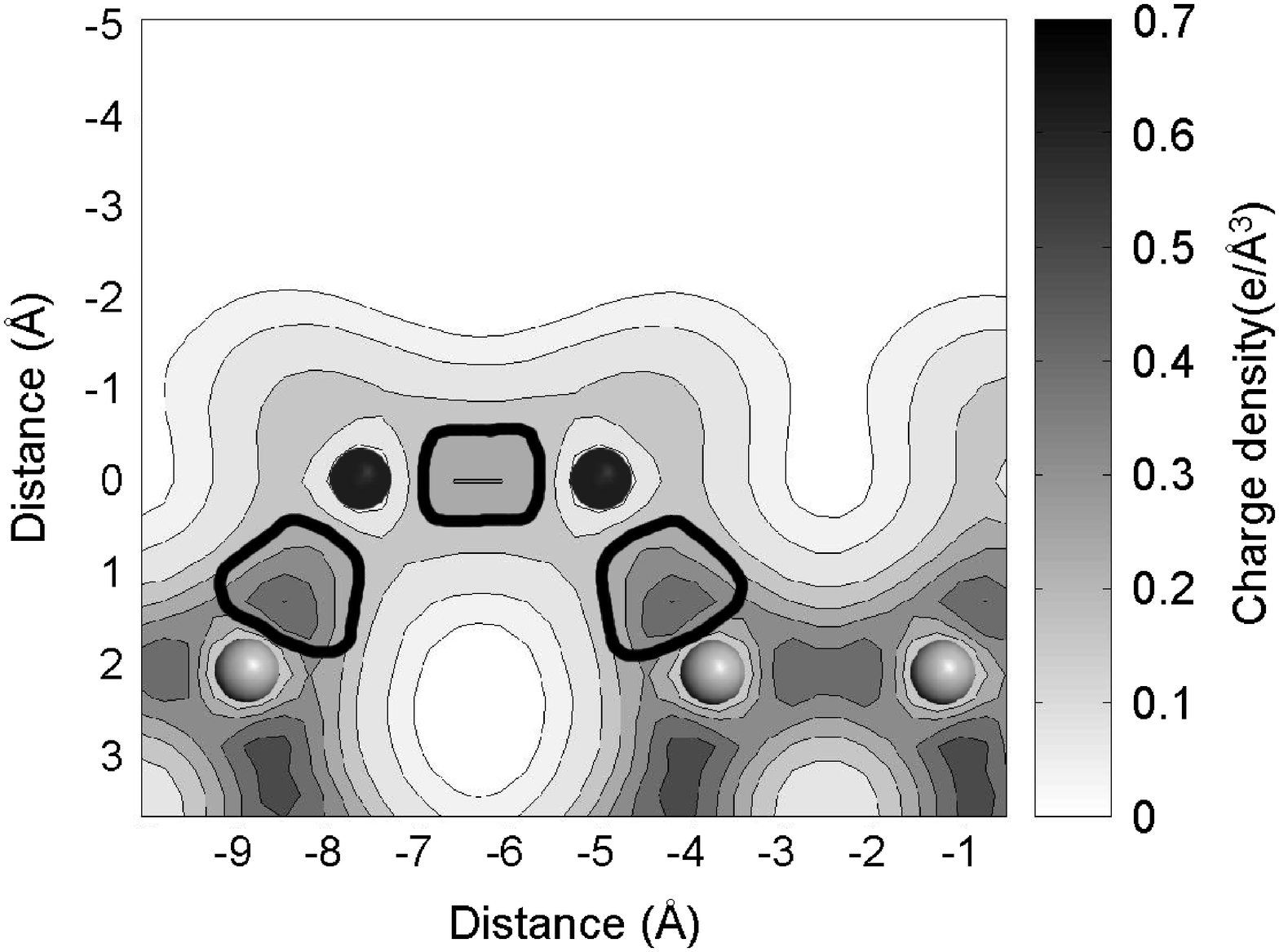}
\caption{}
\label{fig2}
\end{figure}

\begin{figure}\centering
\includegraphics[width=7.cm]{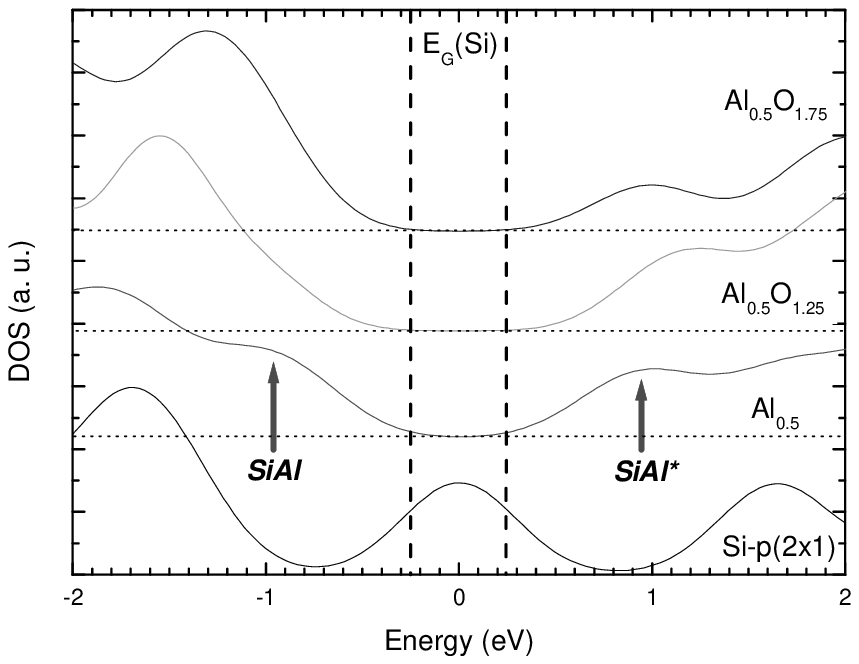}
\caption{}
\label{fig3}
\end{figure}

We carried out electronic structure calculations of this Al-lines template to get insights on the electron distribution and bonds at the surface. We considered a supercell made of a 7 Si layer slab whose bottom layer was fixed and saturated by hydrogen atoms, and top surface was covered by the adatoms and a vacuum slab of 14~$\AA$. Charge density maps (Fig.\ref{fig2}) show high density regions between top Si atoms and Al atoms, representing the interface's Si-Al bonds. We also notice another high density region inside each pair of Al atoms, corresponding to Al-Al bonds. Thus, one may understand that a pair of Al provides 4 electrons to 4 Si dangling bonds while the 2 Al share 2 electrons, leading to a coverage of $1/2$~ML. Electronic densities of states (Fig.\ref{fig3}) shows that the half filled state in the band gap of the Si(001)-p(2$\times$1) disappeared in favor of the Si-Al bond creation. Therefore, electron counting model must be "extended" to take into account the dimer formed by the adatoms.

This Si surface covered by half a layer of Al was then oxidized thanks to two approaches: i) by considering electron counting arguments to obtain a neutral interface, ii) by computing with an \textit{ab initio} method the energy of the system with one O atom at various positions (A-F on Fig.\ref{fig1}) in order to define the most stable ones. Let us first count the electrons. In order to construct a neutral interface, one considers that $\it{N}$ O atoms need 2$\it{N}$ electrons. As the (1$\times$1) surface unit cell provides 2.5 electrons (1 electron from the dangling bonds and 1.5 electrons from the Al$_{0.5}$~ML), the interface is neutral with 1.25 O atoms. Thus, the interface stoichiometry should be Al$_{0.5}$O$_{1.25}$. Now, within \textit{ab initio} computations, each of the A-F site's vertical position was relaxed and the final total energy values were compared. A and C sites have similar energies and are more stable than the others by more than 1~eV per supercell. This may be easily understood if one considers the trend of O atoms to go where a high electron density stands (on Si-Al and Al-Al bonds). These stable sites were then fully filled and lead to the addition of 1.25 O ML which is in agreement with a neutral interface. Let us here mention that the addition of one O atom on each Si-Si bridge (B sites) leads to 1.75 O ML which also leads to a neutral interface.

The resulting Al$_{0.5}$O$_{1.25}$ surface was fully relaxed and its electronic structure studied. Si-O bond length is found to be similar to those in SiO$_2$. Al-O distances (from 1.76 to 1.85~\AA) are compatible with those in an AlO$_2$ plane of LaAlO$_3$. Atoms in the Al$_{0.5}$O$_{1.25}$ layer are nearly in the same plane. Concerning the electronic density of states (Fig.\ref{fig3}), the oxidation does not add states in the gap of the Al covered surface, and the Si gap is finally recovered. The same behaviour occurs for Al$_{0.5}$O$_{1.75}$.

\begin{figure}\centering
\includegraphics[width=7.5cm]{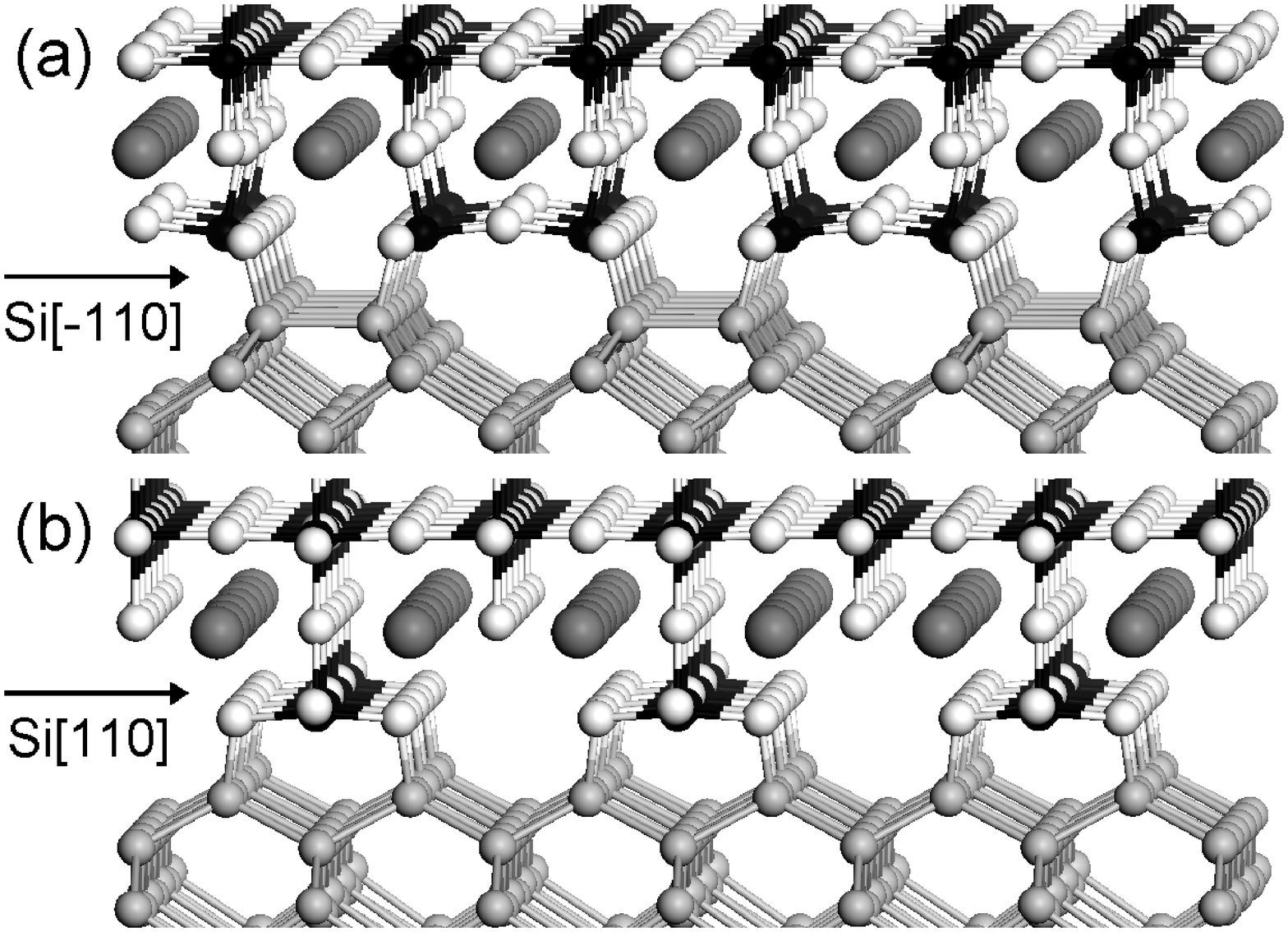}
\caption{}
\label{fig4}
\end{figure}

In order to define further growth over the Al$_{0.5}$O$_{1.25}$ layer, the interface was also capped by a neutral LaAlO$_3$ bilayer and both LaO-AlO$_2$ and AlO$_2$-LaO sequences with various in plane positions were constructed and relaxed (Fig.\ref{fig4}). Their total energies were found to be equivalent, showing that further growth on the interface would be similar whatever its structure would be. Moreover, the perovskite structure of the relaxed bilayer was maintained, demonstrating the possibility of growing LaAlO$_3$ over this interface.

\begin{figure}\centering
\includegraphics[width=8.cm]{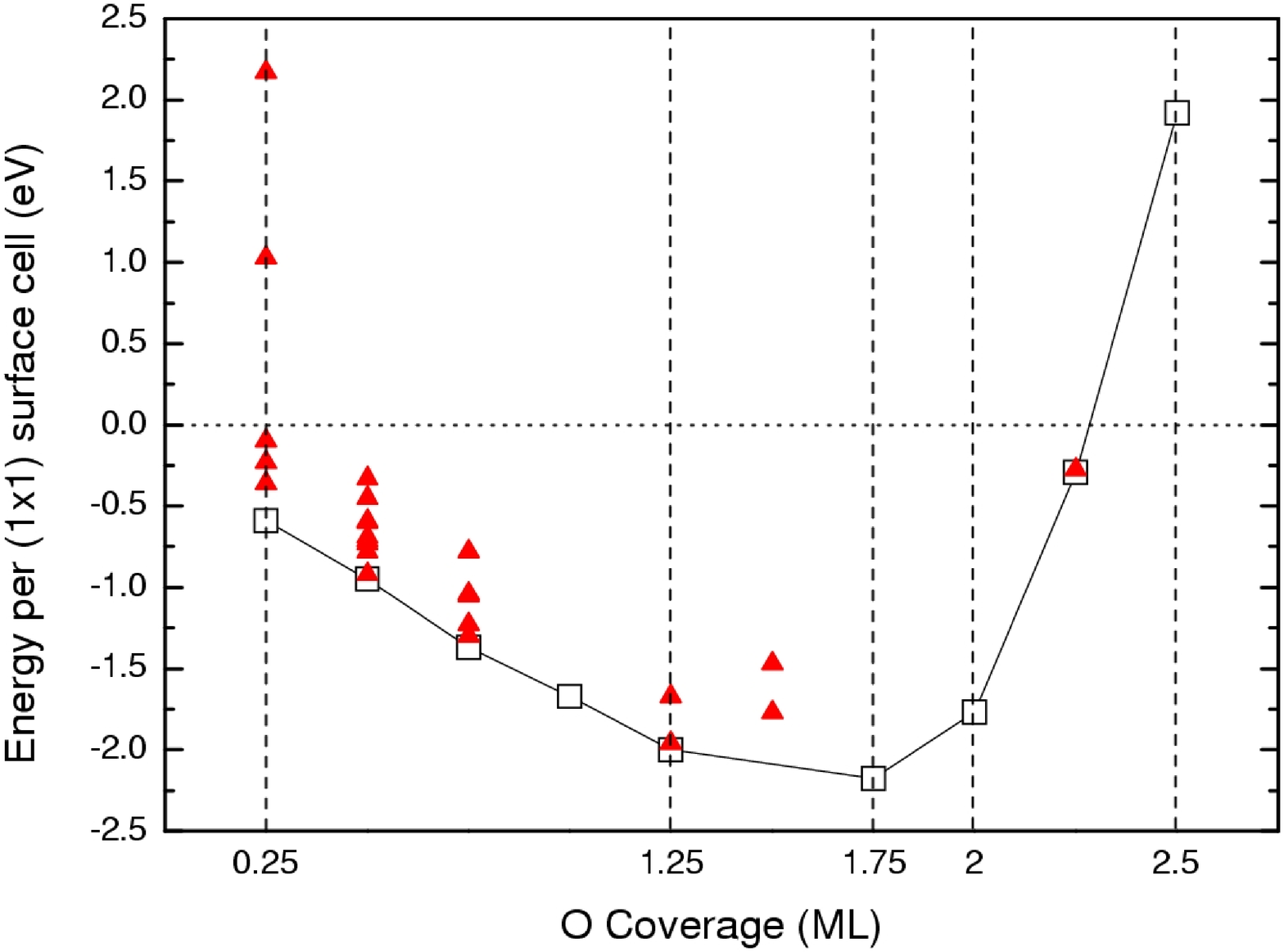}
\caption{}
\label{fig5}
\end{figure}

\begin{figure}\centering
\includegraphics[width=8.cm]{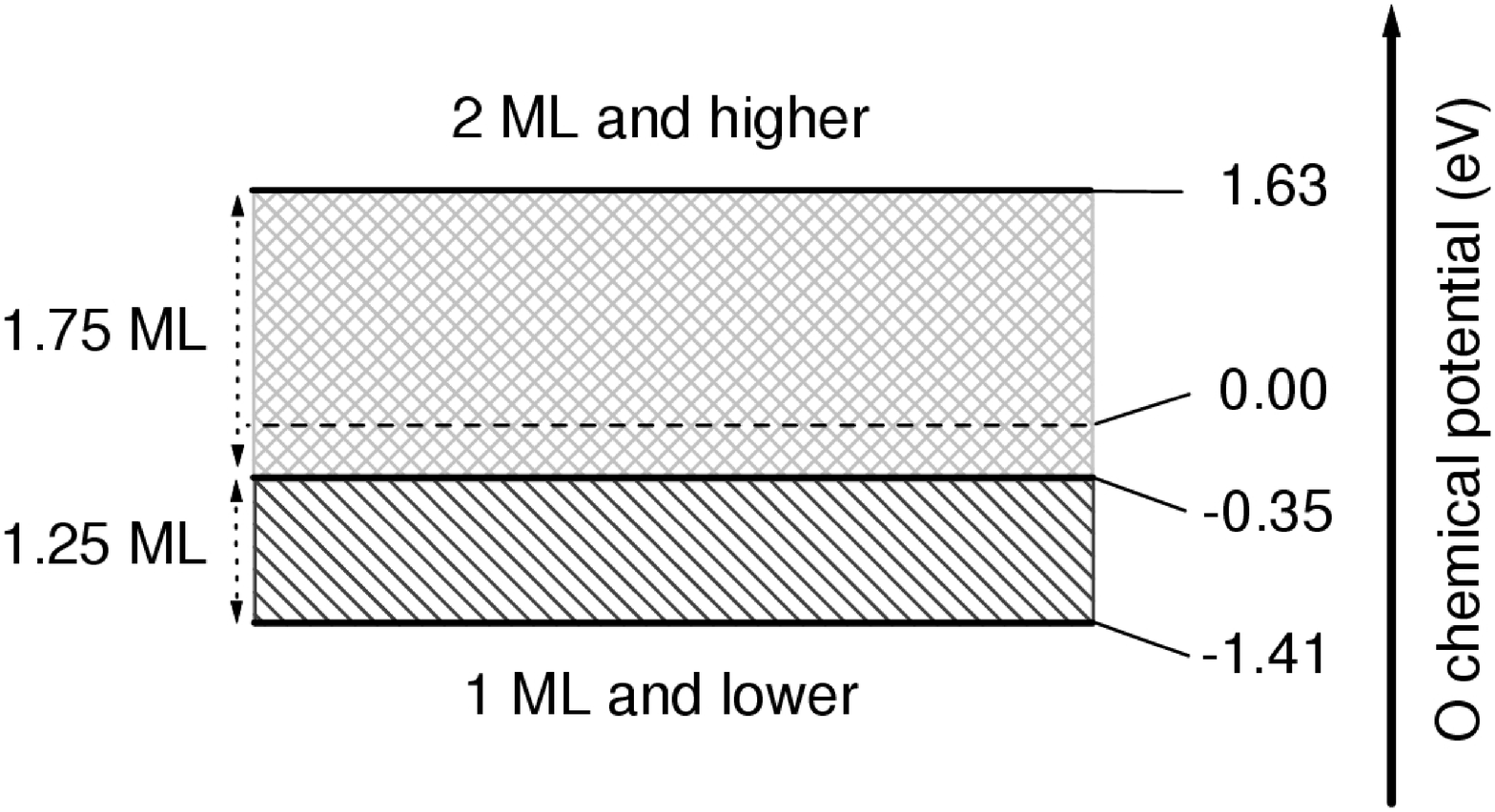}
\caption{}
\label{fig6}
\end{figure}

The stability of this interface against oxidation was tested using the method proposed by F\" orst et al.\cite{ForstStability}. Figure \ref{fig5} shows the adsorption energy of O atoms per (1$\times$1) surface unit cell for various O coverages. When possible, several structures were constructed for a fixed coverage, whose lowest energy corresponds to the most stable one. On this basis, our Al$_{0.5}$O$_{1.25}$ interfacial layer is more stable than other lower coverages and coverages higher than 1.75 (which corresponds to Al$_{0.5}$O$_{1.25}$ with Si-Si oxidized dimers). Using this figure, the oxygen chemical potential was evaluated to construct the phase diagram for interface oxidation (Fig.\ref{fig6}). It shows that: i) for a chemical potential below -0.35~eV, the  Al$_{0.5}$O$_{1.25}$ interfacial layer can be formed without silica, ii) for a chemical potential between -0.35~eV and 0~eV, the Si-Si dimers below the Al$_{0.5}$O$_{1.25}$ interfacial layer are oxidized, leading to the Al$_{0.5}$O$_{1.75}$ stoichiometry.

Concerning applications, a key point is the conduction band offset with Si, which is whished to be greater than 1~eV to avoid tunnelling through the oxide barrier. As the interface between the oxide and Si greatly influences its value, it is important to evaluate the offset for each interface. We computed this offset for the Al$_{0.5}$O$_{1.25}$ interface followed by the previous sequences. The computations were realized on supercells built as follow: H - Si slice (13~$\AA$) - LaAlO$_3$ (19~$\AA$) - Si slice (13~$\AA$) - H containing a mirror plane in order to avoid infinite addition of the interface's electronic dipole. The average potential method was used to evaluate the band offsets \cite{VanDeWalle}. The computed conduction band offsets is 1.9~eV which is large enough for microelectronics applications. In case of Al$_{0.5}$O$_{1.75}$, the offset does not change by more than 0.1~eV.

In conclusion, we have derived an interface between a polar high-$\kappa$ perovskite and Si. The interface corresponds to the prior growth of metal lines followed by oxidation. We emphasize that the Al$_{0.5}$O$_{1.25}$ interface (with or without oxidized Si-Si dimers) leads to consider LaAlO$_3$ as a promising material to replace SiO$_2$ with a conduction band offset of about 1.9~eV. More generally, this result plays in favor of interfaces based on B metal atoms in case of growth of the polar ABO$_3$ perovskite on the Si(001)-p(2$\times$1) surface.

The authors thank Knizhnik et al. for giving their pseudopotentials and basis sets to use in SIESTA.
%
%

\newpage

\noindent Figure \ref{fig1}. Top (a) and side (b) view of the Al lines geometry. The atomic positions are those corresponding to the total energy minimum as found by our computations. The grey bars link Si dimers and the black bars link Al dimers forming the lines. Letters A to F refer to the in plane positions of O atoms that are added to the lines.
\newline

\noindent Figure \ref{fig2}. Charge density map of the Al lines showing the high density areas (bold lines) along the Al-Si and Al-Al bonds (Si and Al atoms are respectively represented by grey and black balls).
\newline

\noindent Figure \ref{fig3}. Electonic densities of states (DOS) around the theoretical Si band gap, $E_G(Si)$, for various coverages on the Si(001)-p(2$\times$1) surface (from the bottom to the top: free Si surface, Al$_{0.5}$, Al$_{0.5}$O$_{1.25}$ and Al$_{0.5}$O$_{1.75}$). The Fermi level stands at 0~eV. SiAl and SiAl$^*$ respectively indicate Si-Al bonding and anti-bonding states.
\newline

\noindent Figure \ref{fig4}. Geometry of the relaxed supercell. Views in bulk silicon's [-110] (a) and [110] (b) directions. White, black, tiny and wide grey balls correspond respectively to O, Al, Si and La atoms.
\newline

\noindent Figure \ref{fig5}. The adsorption energy per (1$\times$1) unit cell as a function of O coverage. Open squares represent thermodynamically accessible structures while triangles correspond to metastable structures.
\newline

\noindent Figure \ref{fig6}. One-dimensional phase diagram of the Al-line interface as a function of the O chemical potential. The dashed line at a chemical potential of 0~eV corresponds to the co-existence of Si and SiO$_2$ ($\alpha$-quartz).

\newpage

\end{document}